\documentclass{article}
\usepackage{amssymb,latexsym}
\usepackage[dvipdf]{epsfig}
\usepackage{color}
\begin{document}
\def\be{\begin{equation}}
\def\ee#1{\label{#1}\end{equation}}

\title{Dark energy interacting with neutrinos and dark matter: a phenomenological theory
}

\author{G. M. Kremer\footnote{kremer@fisica.ufpr.br}\\
Departamento de F\'\i sica, Universidade Federal do Paran\'a\\ Caixa
Postal 19044, 81531-990 Curitiba, Brazil}

\maketitle
\begin{abstract}
A model for a flat homogeneous and isotropic Universe composed of
dark energy, dark matter, neutrinos, radiation and baryons is
analyzed. The fields of dark matter and neutrinos are supposed to
interact with the dark energy. The dark energy is considered to obey
either the
 van der Waals or the Chaplygin equations of state. The ratio between the
 pressure and the energy density
of the neutrinos varies with the red-shift simulating  massive and
non-relativistic neutrinos at small red-shifts and non-massive
relativistic neutrinos at high red-shifts. The model can
reproduce the expected red-shift behaviors of the deceleration
parameter and of the density parameters of each constituent.

\end{abstract}
%\pacs{98.80.-k, 95.36.+x, 95.35.+d }

The recent astronomical measurements of type-IA
supernovae~\cite{1,2,3,4} and the analysis of the power spectrum of
the CMBR~\cite{5,6,7,8,9} provided strong evidence for a present
accelerated expansion of the Universe~\cite{3,10,11,13,14,15}; the
nature of the responsible entity, called dark energy, still remains
unknown. Furthermore, the measurements of the rotation curves of
spiral galaxies \cite{16} as well as other astronomical experiments
suggest that the luminous matter represents only a small amount of
the massive particles of the Universe, and that the more significant
amount is related to dark matter. That offered a new setting for
cosmological models with dark energy and dark matter and in these
contexts
 many interesting phenomenological models
appear in the literature analyzing  the interaction of
neutrinos~\cite{17,18,19} and dark matter~\cite{20,21,22,23,24,25}
with dark energy. With respect to dark energy some exotic equations
of state were proposed in the literature and among others we quote
the van der Waals~\cite{26,27,28,29,30} and the
Chaplygin~\cite{31,32,33,34} equations of state.

In the present work a very simple cosmological model -- for a homogeneous,
isotropic and flat Universe composed by dark matter, dark energy, baryons, radiation
and neutrinos -- is investigated where the
 dark energy is modeled either by the  van der Waals or the Chaplygin
equations of state and interact with neutrinos and dark matter.
Units have been chosen so that $8 \pi G/3 = c = 1$, whereas the
metric tensor has signature ($+,-,-,-$).

Let a homogeneous, isotropic and spatially flat Universe be
characterized by the Robertson Walker metric
$ds^2=dt^2-a(t)^2\delta_{ij}dx^idx^j$, where $a(t)$ denotes the
cosmic scale factor. The sources of the gravitational field are
related to a mixture of five constituents described by the fields of
dark energy, dark matter, baryons, neutrinos and radiation. The
components of the energy-momentum tensor of the sources is written as
 \be
 ({T^\mu}_\nu)={\rm diag}(\rho,-p,-p,-p),
 \ee{1}
 where $\rho$ and $p$ denote the total energy density and pressure of the sources,
 respectively. In terms of the energy densities and pressures of the constituents
 it follows
 \be
 \rho=\rho_{ b}+\rho_{ dm}+\rho_{ r}+\rho_\nu+\rho_{ de},\quad
 p=p_{ b}+p_{ dm}+p_{ r}+p_\nu+p_{ de}.
 \ee{2}
 Above the indexes ($ b$, $ dm$, $ r$, $\nu$, $ de$) refer to the
 baryons, dark matter, radiation, neutrinos and dark energy, respectively.

The conservation law of the energy-momentum tensor
${T^{\mu\nu}}_{;\nu}=0$ leads to the evolution equation for the
total energy density of the sources, namely
 \be
 \dot\rho+3{\dot a\over a}(\rho+p)=0,
 \ee{3}
 where the dot refers to a differentiation with respect to time.

 The baryons and radiation are considered as non-interacting
 fields so that the evolution equations for their energy densities read
  \be
 \dot\rho_{ b}+3{\dot a\over a}\rho_{ b}=0,\qquad
 \dot\rho_{ r}+4{\dot a\over a}\rho_{ r}=0,
 \ee{4}
 once the baryons represent a pressureless fluid, i.e.,
 $p_{ b}=0$,  and the radiation pressure is given in terms of
 its energy density by $p_{ r}=\rho_{ r}/3$.

 According to a  model proposed by Wetterich~\cite{20}
 the evolution equation for the energy density of a pressureless $(p_{dm}=0)$  dark matter field
 which interacts with a scalar field $\phi$ is given by
 \be
 \dot\rho_{dm}+3{\dot a\over a}\rho_{dm}=\beta\rho_{dm}\dot\phi.
 \ee{5}
 Here the scalar field plays the role of the dark energy and  $\beta$ is a constant
 which couples the fields of dark matter and dark energy.

 For interacting neutrinos with dark energy it is supposed that the evolution equation of the energy density
 is given by (see~\cite{18,19})
 \be
 \dot\rho_\nu+3{\dot a\over
 a}(\rho_\nu+p_\nu)=\alpha(\rho_\nu-3p_\nu)\dot\phi.
 \ee{6}
 The coefficient   $\alpha$ is connected with the mass of the
 neutrinos and for more details one is referred to~\cite{18,19} and to the references therein.
 Here $\alpha$ will be consider a phenomenological
 coefficient that couples the dark energy field with the neutrinos.
 Note that if $p_\nu=\rho_\nu/3$, there is no coupling between the
 fields of dark energy and neutrinos. Moreover, it is also important to  note that the neutrinos
 in the past must behave as massless particles where the relationship
 between the pressure and the energy density is $p_\nu=\rho_\nu/3$. Due
 to the coupling of the neutrinos with the scalar field they become
 massive and non-relativistic. For these reasons a barotropic
 equation of state for the neutrinos is proposed where the ratio
 between the pressure and the energy density $w_\nu=p_\nu/\rho_\nu$,
 given in terms of the red-shift $z$, reads
 \be
 w_\nu=\left[{1\over z^2}+{5\over z}{K_3(1/z)\over K_2(1/z)}-{1\over
 z^2}\left({K_3(1/z)\over K_2(1/z)}\right)^2-1\right]^{-1}.
 \ee{7}
Above $K_2(1/z)$ and $K_3(1/z)$ are modified Bessel functions of
second kind. For small values of $z$, $w_\nu$ tends to the
non-relativistic limit equal to 2/3, whereas for large values of
$z$, $w_\nu$ tends to the relativistic limit equal to 1/3. It is
noteworthy that for red-shifts $z\approx10$ this ratio reaches the
value  $w_\nu\approx1/3$ and the coupling between the neutrinos and
the dark energy is negligible. The expression given in (\ref{7}) is
motivated by the equation of the specific heat of a relativistic gas
(see e.g.~\cite{35}).

 The evolution equation for the energy density of the dark energy
 field is obtained from equations (\ref{2}) through (\ref{6}),
 yielding
 \be
 \dot\rho_{de}+3{\dot a\over
 a}(\rho_{de}+p_{de})=-\alpha\dot\phi(\rho_\nu-3p_\nu)-\beta\rho_{dm}\dot\phi.
 \ee{8}
 The energy density and pressure of the dark energy are connected
 with the scalar field by $\dot\phi=\sqrt{\rho_{de}+p_{de}}$. Since
 the purpose of this work is to develop a phenomenological theory,
 it  is assumed that the dark energy field behaves either as a van der
 Waals or a Chaplygin fluid with an equation of state given by~\cite{29,30,31,32,33,34}
 \be
 p_{vw}={8w_{vw}\rho_{vw}\over3-\rho_{vw}}-3\rho_{vw}^2,\qquad p_{ch}=-{A\over\rho_{ch}},
 \ee{9}
 where $w_{vw}$ and $A$ are positive free parameters in the van der Waals and Chaplygin
 equations of state, respectively.

 For the determination of the time evolution of the energy densities one
 has to close the system of differential equations by introducing the Friedmann equation
 \be
 \left({\dot a\over a}\right)^2=\rho.
 \ee{10}

 From now on the red-shift will be used as a variable instead of time
 thanks to the following relationships
 \be
  z={1\over a}-1,\qquad {d\over dt}=-\sqrt{\rho}(1+z){d\over dz}.
  \ee{11}

 Equations (\ref{4}) can be easily integrated leading to the
 well-known dependence of the energy densities of the baryons and
 radiation with the red-shift
 \be
 \rho_r(z)=\rho_r(0)(1+z)^4,\qquad
 \rho_b(z)=\rho_b(0)(1+z)^3,
 \ee{12}
 whereas equations (\ref{5}), (\ref{6}) and (\ref{8}) become a system
 of coupled differential equations for the energy densities
 $\rho_{dm}$, $\rho_\nu$ and $\rho_{de}$, namely,
 \be
 {(1+z)\rho_{dm}^\prime-3\rho_{dm}\over\sqrt{(\rho_{de}+p_{de})/\rho}}
 =-\beta\rho_{dm},
 \ee{13}
 \be
 {(1+z)\rho_\nu^\prime-3(\rho_\nu+p_\nu)\over\sqrt{(\rho_{de}+p_{de})/\rho}}
 =-\alpha(\rho_\nu-3p_\nu),
 \ee{14}
 \be
 {(1+z)\rho_{de}^\prime-3(\rho_{de}+p_{de})\over\sqrt{(\rho_{de}+p_{de})/\rho}}=\beta\rho_{dm}
  +\alpha(\rho_\nu-3p_\nu).
 \ee{15}
 In the above equations the prime refers to a differentiation with
 respect to the red-shift.

\begin{figure}%\vskip0.1cm
\begin{center}
\includegraphics[width=7.5cm]{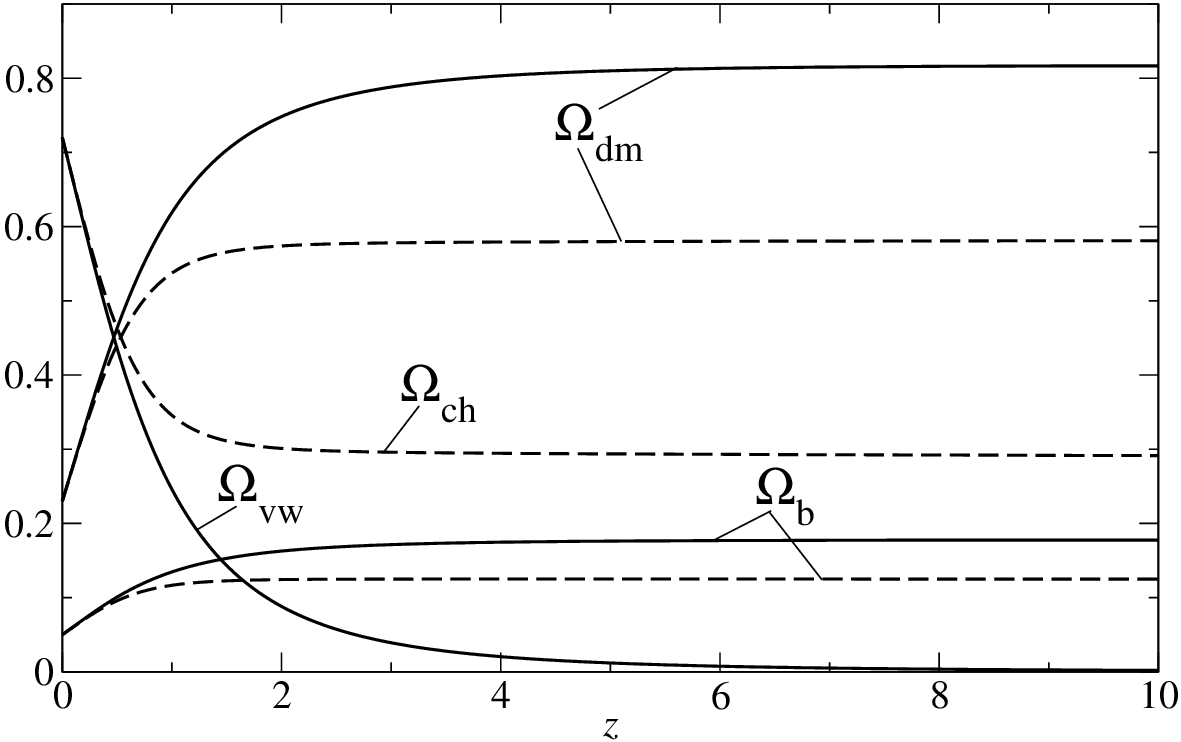}%\vskip0.1cm
\caption{Density parameters as functions of red-shift: van der Waals fluid (solid lines)
and Chaplygin fluid (dashed lines). }
\end{center}
\end{figure}

In order to solve the coupled system of differential equations
(\ref{13}) -- (\ref{15}) one has to specify initial values for the
energy densities at $z=0$. The following initial values for the
density parameters $\Omega_i(z)=\rho_i(z)/\rho(z)$ taken from the
literature (see~\cite{36} for a review) were chosen:
 $ \Omega_{de}(0)=0.72,$ $\Omega_{dm}(0)=0.229916,$
 $\Omega_b(0)=5\times10^{-2},$
 $\Omega_{r}(0)=5\times10^{-5},$ $\Omega_\nu(0)=3.4\times10^{-5}$.
 Moreover, one has to specify values for the coupling parameters $\alpha$
 and $\beta$ and for the parameters $w_{vw}$ and $A$ which appear in the van der Waals and
 Chaplygin equations of state (\ref{9}). One way to fix the two last parameters
 is through the use of the value of the deceleration parameter
 $q=1/2+3p/2\rho$ at $z=0$. Indeed, by considering $q(0)=-0.55$ it
 follows $w_{vw}=0.33851$ and $A=0.50403$. For the coupling parameters two sets of values
 were chosen, namely,  (a) $\alpha = 5\times10^{-5}$ and $\beta = -5\times10^{-5}$
 for the van der Waals equation of state and (b) $\alpha = 10^{-1}$ and $\beta = -10^{-2}$
 for the Chaplygin equation of state. Its is also important to note that by increasing the
 value of the coupling parameter $\alpha$ (and/ or $\beta$)
 the transfer of energy between the dark energy and neutrinos (and/or dark matter) becomes more efficient.

\begin{figure}
\begin{center}
\includegraphics[width=7.5cm]{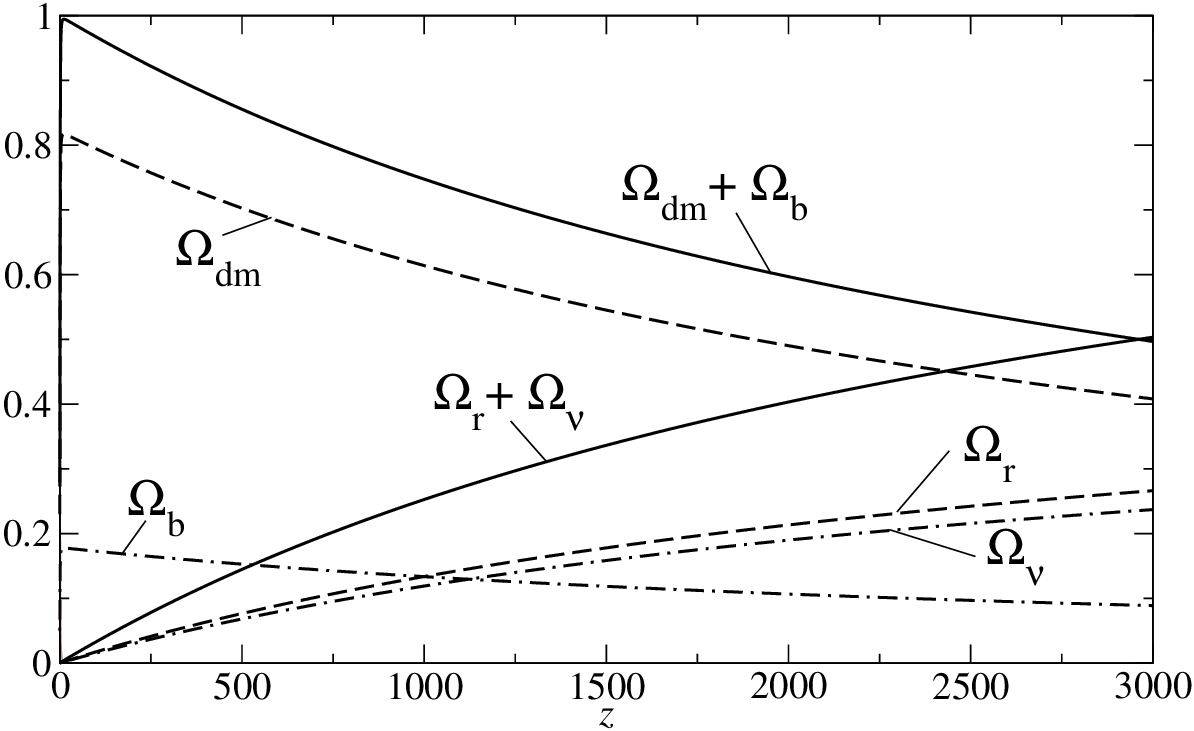}
\caption{Density parameters as functions of red-shift for a van der
Waals fluid as dark matter.}
\end{center}
\end{figure}

In Fig. 1 the density parameters are plotted as functions of the red-shift for  values in the range
$0\leq z\leq10$. The straight lines refer to the case where the van der Waals equation of state is used to describe
the dark energy field whereas the dashed lines correspond to the Chaplygin equation of state. The two density
parameters that represent the dark energy field are denoted  by $\Omega_{vw}$ and $\Omega_{ch}$.
One can infer from this figure that the dark energy density parameter tends to zero for high red-shifts
when the van der Waals equation of state  is used, whereas it tends to a
constant value for the Chaplygin equation of state.
While for high red-shifts the van der Waals equation of state simulates a cosmological constant with
$p_{vw}=-\rho_{vw}$, the pressure of the Chaplygin fluid vanishes indicating that it becomes another
component of the dark matter field (see also the behavior of the pressures indicated in Fig. 4). It is also
important to note that the density parameters of the baryons and of the dark matter increase more with the red-shift
for the van der Waals equation of state, since there is an accentuated decrease in the density parameter of the dark
energy for this case. Note that the density parameters of the radiation and neutrinos are very small in this range of
the red-shift and are not represented in this figure.

The behavior of the density parameters for the cases of the van der
Waals  and Chaplygin equations
 of state are shown in Figs. 2 and 3, respectively,  for red-shifts in the range $0\leq z \leq3000$.
 One can conclude from these figures, as expected, that the density parameters of the neutrinos and
 radiation increase with the red-shift whereas those of the baryons and dark matter decrease.
 Furthermore, the equality between the ``matter"  and ``radiation" fields occurs when $z\approx 3000$ for the
 case where the dark matter field is modeled as a van der Waals  fluid and $z\approx4200$ for the case
 of a Chaplygin fluid. This can be easily understood, since in the latter case the dark energy becomes dark matter
 for high red-shifts contributing for the density parameter of the ``matter" field.
\begin{figure}
\begin{center}
\includegraphics[width=7.5cm]{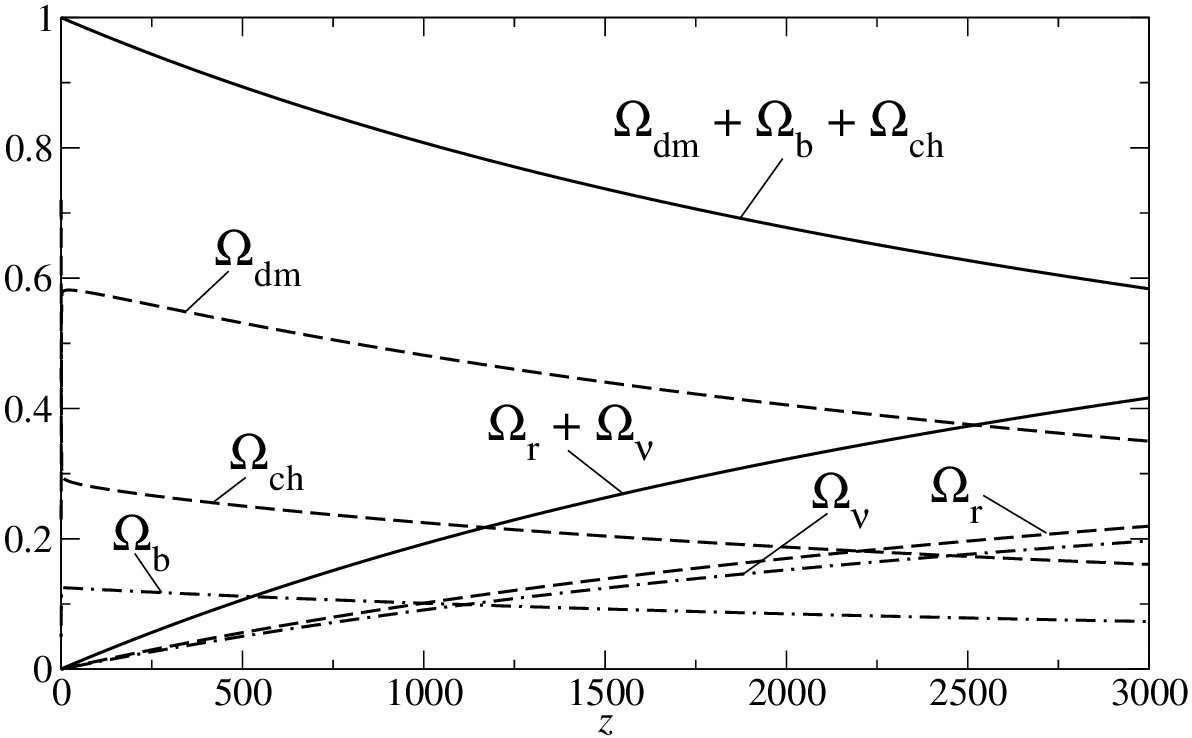}
\caption{Density parameters as functions of red-shift for a
Chaplygin fluid as dark matter.}
\end{center}
\end{figure}
\begin{figure}
\begin{center}
\includegraphics[width=7.5cm]{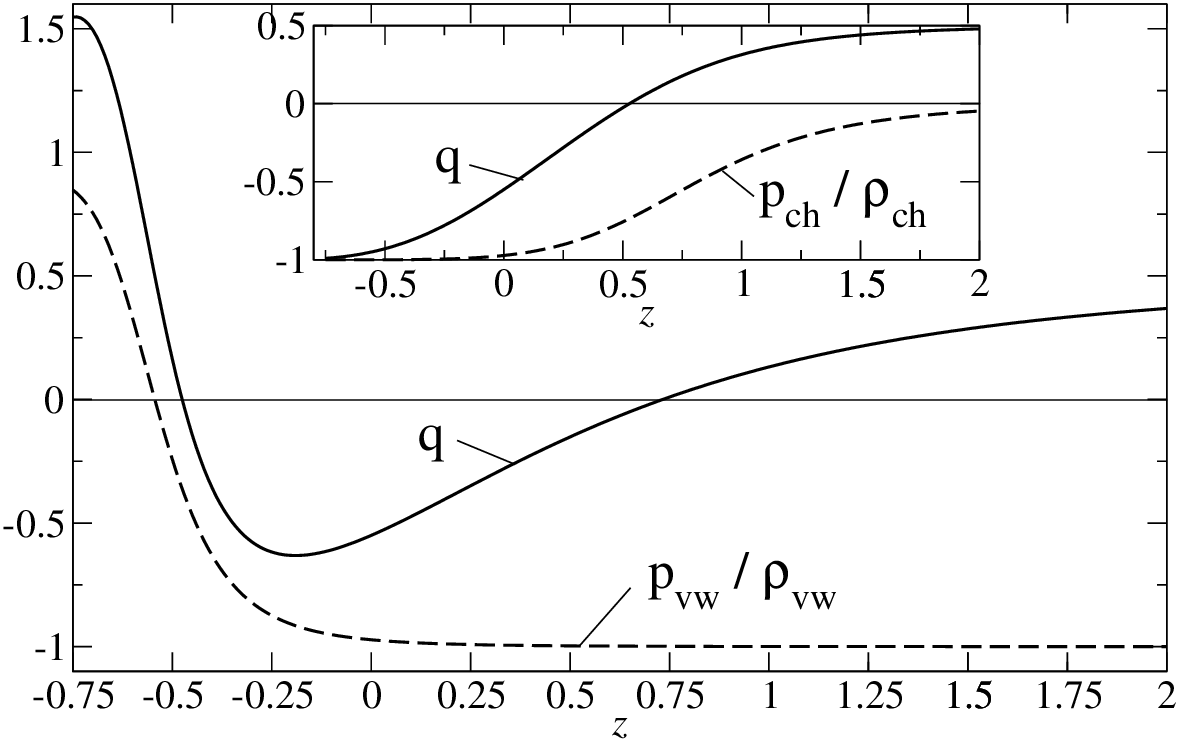}
\caption{Deceleration parameter and ratio between the pressure and
the energy density as functions of red-shift: large frame (van der
Waals), small frame (Chaplygin).}
\end{center}
\end{figure}

In Fig. 4 are plotted the deceleration parameter and the ratio between the pressure
and the  energy density for both cases, the large frame corresponding
to the van de Waals fluid whereas the small frame to the Chaplygin fluid.
 For both cases the deceleration parameter at $z=0$ is equal to $q(0)=-0.55$, since this value was fixed in order to
 find the parameters $w_{vw}$ and $A$ in the equations of state (\ref{9}). The transition from the decelerated to the
 accelerated phase of the Universe occurs at $z_T=0.73$ and $z_T=0.53$ for the van der Waals and
 Chaplygin equations of state, respectively. It is interesting to note that while the Chaplygin
 equation of state simulates a cosmological constant with $p_{ch}=-\rho_{ch}$ for
 negative red-shifts which implies an accelerated phase of the Universe in the future,
 the van der Waals equation of state leads to
 a positive pressure and brings  the Universe to another decelerated phase in the
 future. It is noteworthy to call attention that for positive values of the
red-shift, the solution of the coupled differential equations
(\ref{13}) through (\ref{15}) predicts that the van der Waals fluid
behaves close to a cosmological constant with
$p_{vw}\approx-\rho_{vw}$. This behavior does not lead to a new
transition from a decelerated to an accelerated phase in the very
early Universe, since the energy density of the radiation field
increases so that the radiation pressure  becomes larger than that
of the van der Waals fluid. For high red-shifts the Universe first
becomes dominated by the baryon and dark matter fields and for
higher red-shifts by the radiation field. This model does not
attempt to model the inflationary period, where the inflaton field
dominates a short rapid evolution of the Universe.

As final remarks we call attention to  the fact that one expects
that the coupling between  dark energy, dark matter and neutrinos
should be weak so that the parameters $\alpha$ and $\beta$ are
restricted to small values. The difference between the parameters
adopted for the van der Waals and Chaplygin equations of state is
due to stability conditions of the non-linear coupled system of
differential equations (\ref{13}) -- (\ref{15}), the van der Waals
equation of state being more unstable for large values of these
parameters than the Chaplygin equation of state. In Fig. 5 we have
plotted the density parameters as functions of the red-shift for the
case where a Chaplygin equation of state is used as dark energy. One
can infer from this figure that the decay of the dark energy density
parameter and the increase of the dark matter density parameter with
the red-shift are more pronounced when there exists a coupling
between the fields. The density parameter of the baryons remains
unchanged since the baryons are uncoupled.

As final comment it is important to note that even without couplings
between the fields of dark energy, dark matter and neutrinos, this
phenomenological model -- with the equations of state of van der
Waals and Chaplyging as dark energy -- can describe satisfactorily
the evolution of a Universe whose constituents are dark energy, dark
matter, baryons, neutrinos and radiation.

\begin{figure}
\begin{center}
\includegraphics[width=7.5cm]{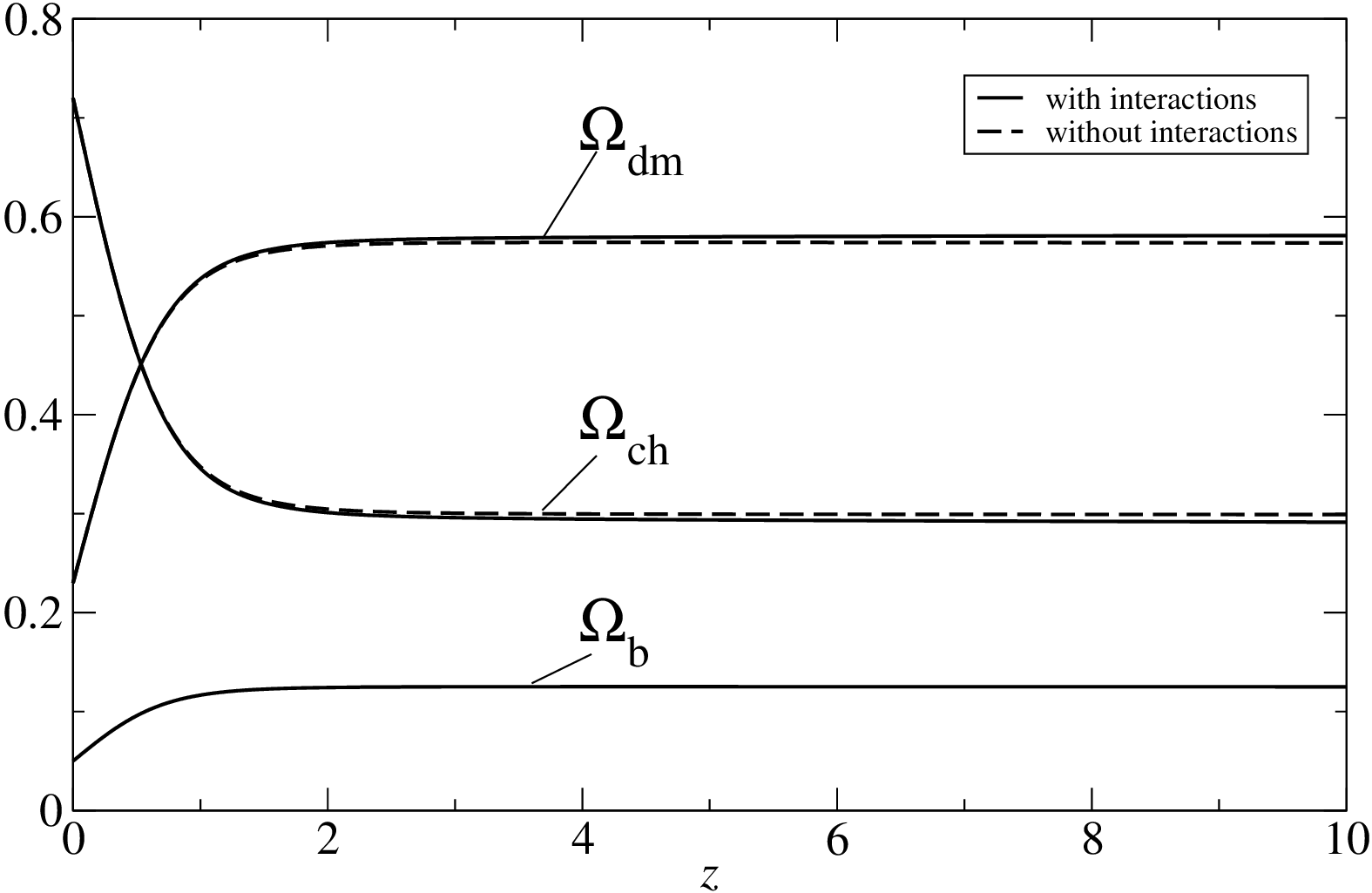}
\caption{Density parameters as functions of red-shift for a
Chaplygin fluid with and without interactions.}
\end{center}
\end{figure}

%%%%%%%%%%%%%%%%%%%%%%%%%%%%%%%%%%%%%%%%%%%%%%%%%%%%%%%%%%%%%%%%%%%%%%%%%%

%%%%%%%%%%%%%%%%%%%%%%%%%%%%%%%%%%%%%%%%%%%%%%%%%%%%%%%%%%%%%%%%%%%%%%%%%%

\end{document}